# ARTICLE

# Atomic scale modeling of water and ice behavior on vibrating surfaces: towards design of Surface Acoustic Wave anti-icing and de-icing systems

Tomasz Wejrzanowski,*[a,b] Stefan Jacob,[c] Andreas Winkler,[d] Agustín R. González-Elipe,[e] Ana Borrás [e]

Within these studies, atomic scale molecular dynamics simulations have been performed to analyze the behavior of water droplets and ice clusters on hydrophilic and hydrophobic surfaces subjected to high-frequency vibrations. The methodology applied herewith aimed at understanding the phenomena governing the anti-icing and de-icing process enabled by Surface Acoustic Waves (SAWs). The complex wave propagation was simplified by in-plane and out-of-plane substrate vibrations, which are relevant to individual longitudinal and transverse components of SAWs. Since the efficiency of such an active system depends on the energy transfer from the vibrating substrate to water or ice, the agents influencing such transfer as well as the accompanied phenomena were studied in detail. Apart from the polarization of the substrate vibrations (in-plane/out-of-plane), the amplitude and frequency of these vibrations were analyzed through atomic scale modeling. Further, the surface wettability effect was introduced as a critical factor within the simulation of water or ice sitting on the vibrating substrate. The results of these studies allow identifying the different phenomena responsible for water and ice removal from vibrating surfaces depending on the wave amplitude and frequency. The importance of substrate wetting for the ant-icing and de-icing has also been analyzed and discussed concerning the future design and optimization of SAW-based systems.

## Introduction

Water and/or ice interaction with materials surfaces is of key importance both for life sciences[1,2] and materials and engineering.[3,4] The hydrophilicity of most natural material surfaces enables water to be transported over large distances via capillary effects.[5,6] Similarly, some plants and insects use a particular design of their outer surface to promote water condensation.[7] Inspired by these observations from nature, specifically designed engineered materials have been proposed to achieve efficient control of wettability. In general, this engineering involves surface treatments aiming at changing both the chemistry and topography of surfaces. Control of surface nanostructure and chemistry is particularly critical to obtain superhydrophobic surfaces (i.e., water contact angle >150°)[8–11], as illustrated in literature by the application of coatings and surface treatments endowing almost full water repellency.[4,12,13] Experimental studies and numerical simulations have been profusely applied to understand the wetting behavior of surfaces (for example in terms of the Wenzel and Cassie-Baxter states) as a function of either intrinsic material properties such as surface topography and chemistry or external conditions (temperature and pressure).[14–21]

The icing of materials is a critical issue for many processes and engineering structures, for example, aircrafts,[22,23] wind turbines,[24,25] or marine structures.[26,27] Trying to circumvent the problems associated with the icing of surfaces, both passive (i.e., preventing the formation of ice) and active (i.e., provoking the removal of ice) solutions have been proposed. The passive or anti-icing solutions have been widely studied during the last years, having found a certain, but not complete correlation between superhydrophobicity and highly efficient anti-icing response, this latter in terms of prevention of ice formation and accretion.[28,29] In fact, to obtain a high anti-icing performance, the material surfaces should also hold other important features such as a high capability for delaying freezing or a low ice adhesion.[30] Regarding de-icing, a general operational condition is the availability to continuously deliver energy or chemical flow to the surfaces to promote the removal of ice. For example, commercial systems for aircrafts are based on thermal heating,[31,32] pneumatic de-ice boots,[33,34] or the application of de-icing lubricants.[35] Drawbacks of most common de-icing systems are high energy consumption, low flexibility with regard to materials and environmental compatibility, or limited duration of their effects.

Recently, the use of ultrasounds[36,37] and, more specifically, short-range wave Acoustic and Surface Acoustic Waves (SAW)[38,39] have been proposed as efficient alternative solutions for de-icing or to prevent icing (to induce an active anti-icing

[d] *Warsaw University of Technology, Faculty of Materials Science and Engineering, Woloska 141, 02 507 Warsaw, Poland*
*E-mail: tomasz.wejrzanowski@pw.edu.pl*
[b] *Technology Partners Foundation, Pawinskiego 5A, 02-106 Warsaw, Poland*
[c] *Physikalische-Technische Bundesanstalt, Bundesallee 100, 38116 Braunschweig, Germany*
[d] *Leibniz IFW Dresden, SAW Lab Saxony, Helmholtz str. 20, 01069 Dresden, Germany*
[e] *Nanotechnology on Surfaces and Plasma Lab, Materials Science Institute of Seville (CSIC-US), Américo Vespucio 49, 41092, Seville, Spain*

effect) at relatively low energetic costs and high operational flexibility.[40,41] The interaction of a SAW propagating along a material with water on its surface is a quite mature subject of the investigation with even practical engineering solutions in the market for quite a wide range of applications including micro-fluidics[42,43] and lab-on-a-chip systems.[44,45] However, the study of the effects of the SAW with accreted ice is still in its infancy and lacks fundamental knowledge about the interaction of ice with the atomic surface oscillations typical of these waves which might shed critical information on the most efficient solutions out the several possibilities for operational surface and bulk acoustic waves. This and similar processes such as fluid nebulization, surface wetting, or ice melting are fast processes that are initiated locally, and as such difficult to characterize with sufficient precision. In this context, atomic scale simulation becomes a useful tool to simulate water and ice behavior on the surface of materials. It has been used to understand the chemical and physical bonding of water molecules to various surfaces.[46–51] Also freezing delay has been recently analyzed for nanoscale systems,[52] aiming at designing superhydrophobic and icephobic coatings and surfaces. Recently, a few theoretical works have also addressed the water behavior on vibrating substrates activated by SAWs.[53,54] However, to the best of our knowledge, no atomic simulation works exist so far about the interaction of SAW with ice. Previous reports about the interaction of SAWs with water droplets suggested that the interaction is based on the transfer of the kinetic energy associated with the vibrating atoms of the substrate to the water molecules in contact with it. The energy transfer interaction will be modeled in this work assuming the effect of a vibrating rigid substrate of ordered atoms at the interface with either water or ice.

In a standard SAW device, the electromechanical excitation is typically induced by a transducer formed by a comb-shape interdigitated electrode (interdigital transducer, IDT), deposited on a piezoelectric substrate or a piezoelectric thin film. The wave generated in such a way propagates along the surface of the substrate and transfers efficiently the energy to water (or ice) on the surface.[55] Rayleigh waves are the most common SAW mode for actuation and energy transfer into a half space atop. Rayleigh waves present characteristic frequency, wavelength, and direction that depend on IDT architecture and orientation, as well as piezoelectric material properties, and are characterized by a sagittal polarization, i.e. with in-plane (longitudinal) and out-of-plane (surface-normal) components.

This work aims to gain knowledge about the fundamental mechanisms involved in the transfer of energy between a vibrating surface and an ice aggregate placed on its surface. For comparison, a similar approach is applied to a water droplet. The objective is to unravel some of the basic mechanisms governing water and ice removal and the effect of such processes on the wetting state of the material surfaces (i.e., its hydrophobic or hydrophilic character). An additional goal is to determine basic SAW characteristics responsible for water or ice activation, mainly the effect of the in-plane and out-plane components of the wave on the effectiveness of the de-icing and water removing processes. The results show that the effectiveness of water removal and ice melting varies differently with the amplitude, frequency, and horizontal/vertical components of the atomic surface vibrations and, in general, are favored by the hydrophobic character of the surface state.

## Materials and methods

Herein we propose to employ a rigid substrate vibrating either horizontally or vertically as a model system to simulate by molecular dynamics (MD) the effect of atomic surface vibrations typical of SAWs on the excitation of a water droplet or an ice cluster deposited on their surface. We assume that these substrate vibrations and the mechanical interaction occurring with the droplet/ice-cluster placed on their surface reproduce the mechanical energy transfer taking place in real systems between the SAW-activated substrates and the droplets/ice-cluster on their surfaces. Concretely and as pointed out in the introduction, SAW has in-plane and out-of-plane components that are simulated in the paper by the vertical and horizontal vibration modes of the substrate, providing a way to estimate the most efficient wave mode for energy transfer to either water or ice.

Atomic scale simulations of the behavior of water droplet and ice cluster on a vibrating substrate were performed using classical molecular dynamics (MD) calculations implemented in the LAMMPS software. Parallel computations were carried out using a highly performant in-house computer cluster under the MPI protocol.

Simple Newton's equations of motion are solved within the MD simulations:

$$F_i = m_i \frac{dv_i}{dt} = \sum_{i \neq j}^{N} -\nabla U_{ij}(r_{ij}), \quad (1)$$

where $F_i$, $m_i$, and $v_i$ are the force, mass, and velocity of each particular atom (molecule) under consideration. $N$ indicates the number of atoms and $t$ is time. $U$ is defined as the interatomic potential between two atoms $i$ and $j$ separated by $r_{ij}$. In these simulations the Leonard-Jones and Coulomb potentials have been used:

$$U_{ij}(r_{ij}) = 4\lambda\epsilon_{ij}\left[\left(\frac{\sigma_{ij}}{r_{ij}}\right)^{12} - \left(\frac{\sigma_{ij}}{r_{ij}}\right)^{6}\right] + \frac{q_i q_j}{4\pi\epsilon_0 r_{ij}} \quad (2)$$

where $\sigma$ is the distance at which the interatomic potential is zero, $\epsilon$ is the depth of the potential well, $q$ is the charge of the atomic site, and $\epsilon_0$ is the vacuum permittivity. A $\lambda$ parameter allows us to tune the substrate–liquid interaction strength, and hence to define the static contact angle, $\theta$. It has been adjusted from superhydrophobic ($\lambda = 0.15$, $\theta \approx 160°$) to hydrophilic ($\lambda = 1$, $\theta < 20°$). The first term on the right side of equation 2 represents the intermolecular forces, while the second term represents the electrostatic contributions. Polar water molecules are modeled using the rigid four-site TIP4P/2005 model;[56] this consists of one oxygen (O) site, two charged hydrogen (H) sites, and one massless charged (M) site located along the bisector of the hydrogen atoms, at a distance of 0.1546 Å from the oxygen atom.

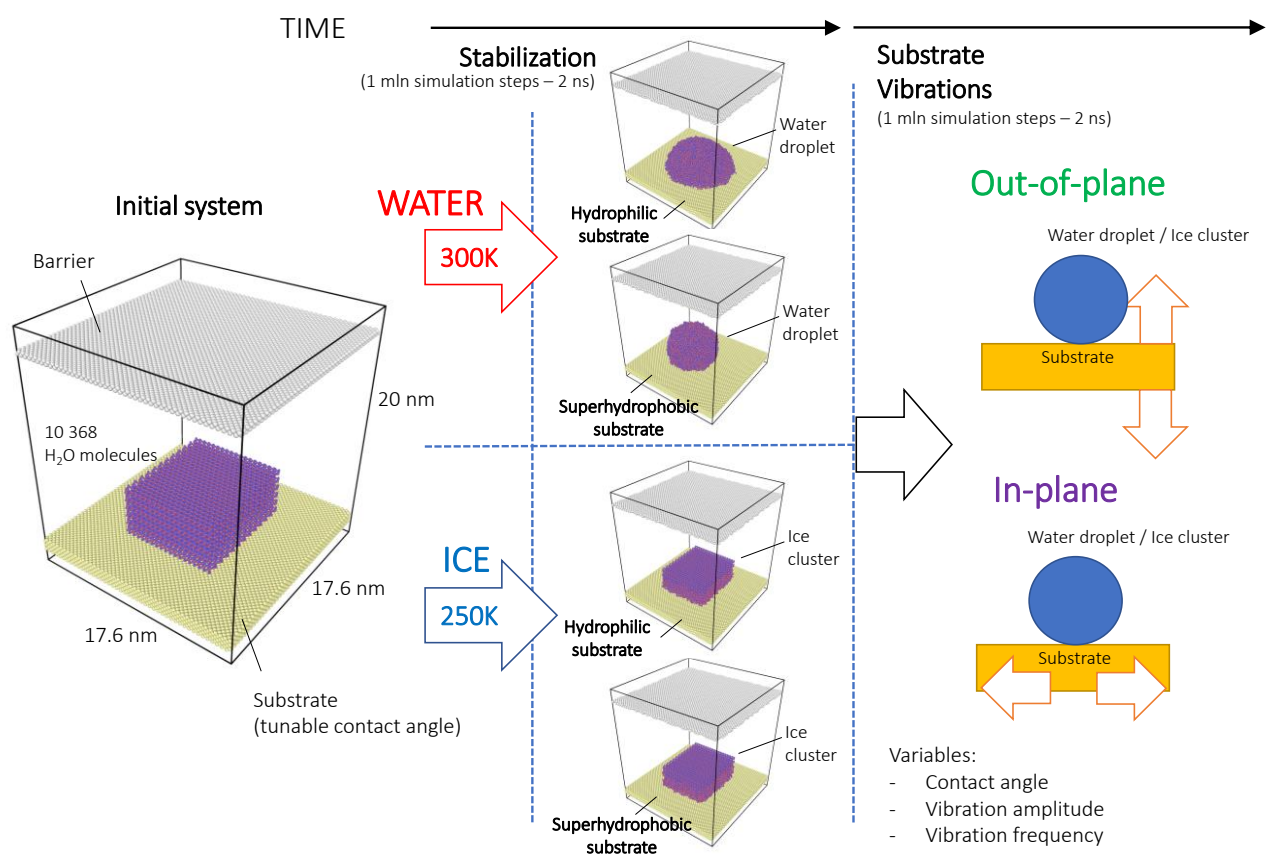

**Fig 1** Schematic illustration of the molecular dynamics simulations from the initial system (left) to the stabilization at two different temperatures (middle) including hydrophilic (WCA≈20°) and superhydrophobic (WCA≈160°) surfaces, and variables for out-of-plane and in-plan vibrations (right).

The internal geometry of the water molecule is constrained by specifying a fixed O−H bond distance (0.9572 Å) and H−O−H angle (104.52°); this structure is maintained using the SHAKE algorithm.[57] The droplet-supporting substrate consists of 6 layers of Pt metal atoms in an FCC lattice, with a lattice constant of 3.92 Å. All the interatomic potential parameters are listed in Table 1.

**Table 1** Interatomic potential parameters and atomic masses ($m_a$) for sites in TIP4P/2005 water atom/molecules (H, O, and M), and substrate Platinum (Pt) atoms.

| Site | $\epsilon$ (kJ/mol) | $\sigma$ (Å) | $q$ (e) | $m$ (u) |
|------|------|------|------|------|
| H | 0 | 0 | 0.5242 | 1.008 |
| O | 0.774 | 3.1589 | 0 | 15.9994 |
| M | 0 | 0 | -1.0484 | 0 |
| Pt | 4.18 | 2.471 | 0 | 195.084 |

As shown in Figure 1, the initial layout used for calculations consists of a cluster of ordered water molecules located on the atomically flat substrate. The domain boundaries are set to be periodic in every direction, and a superhydrophilic barrier is positioned far above the substrate to collect atomized molecules and prevent them from leaving the domain via the top boundary and re-entering through the bottom boundary. The values of σ, ϵ, and atomic mass m for the substrate/barrier atoms are derived from those of platinum, which is selected as the substrate with no water-substrate chemical bonding.

The MD simulation system in Figure 1 comprises a cluster of 10368 water molecules (hexagonal crystal). It is equilibrated for 2 ns at 300 K and 250 K for the stabilization of a water droplet and an ice cluster, respectively. During ice equilibration the outmost top 4 layers were grouped into a rigid body, thus avoiding surface melting effects typical for nanoobjects.[58,59]

Following equilibration, the simulation is run for a period of 2 ns, during which the substrate is oscillated either surface-normal (out-of-plane) or horizontally (in-plane) at frequencies between 50 and 200 GHz, and for a range of amplitudes between 0.1-1 nm. Substrate wettability of 20º (simulating hydrophilicity) and 160º (simulating superhydrophobicity) have been considered for the calculations. Temperature control is not applied to the water molecules during the simulation, but the substrate and barrier molecules are coupled to a Berendsen thermostat to maintain the surface at constant temperature (300 K for water and 250 K for ice cases, respectively).

The OVITO software has been used to visualize the system evolution. To differentiate between ice and water molecules, the oxygen atoms were identified as belonging or not to the hexagonal-type structure characteristic of the initial arrangement of water molecules. This classification tool has been implemented in the OVITO program.[60]

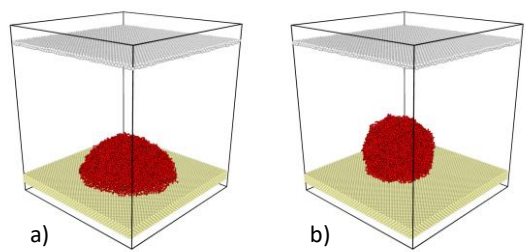

**Fig 2** Atomic structure after 2 ns stabilization for a) hydrophilic and b) hydrophobic surface states.

## Results and discussion

### Vibrational excitation of the water droplet

The initial ice cluster shown in Figure 1 was molten at 300K and the water droplet was stabilized at this temperature for 2 ns. Two different substrates were used in the calculations to simulate the effect of the surface wettability. Atomic structures after 2 ns stabilization at 300K are shown in Figure 2.

After stabilization of the water droplet on the surface, the substrate was subjected to two types of vibrations: horizontal (in-plane) and vertical (out-of-plane). The water droplet behavior was observed for 2ns of simulation time for various values of amplitude and frequency of the substrate vibrations. Specific values of these parameters were varied during the calculations to identify transition limits between various physical phenomena.

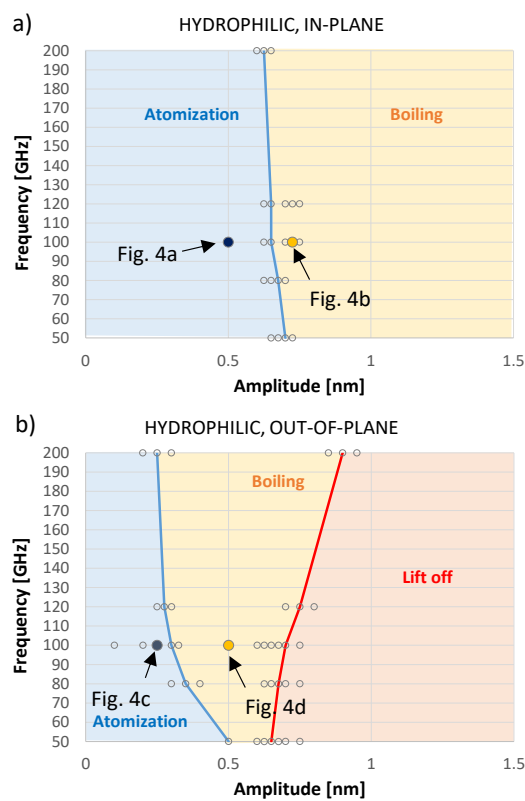

**Fig 3** Maps of excitation phenomena for a water droplet sitting on a hydrophilic substrate subjected to a) in-plane and b) out-of-plane vibrations as a function of specific amplitude and frequency values.

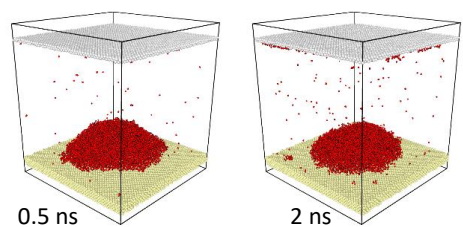

a) Hydrophilic, **in-plane**: 100 GHz, **0.5 nm**

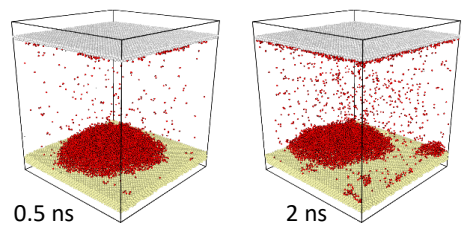

b) Hydrophilic, **in-plane**: 100 GHz, **0.75 nm**

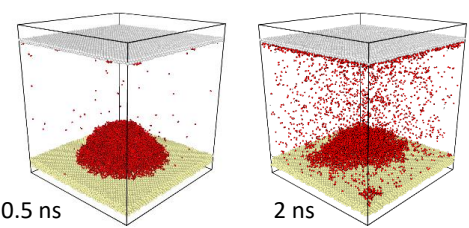

c) Hydrophilic, **out-of-plane**: 100 GHz, **0.25 nm**

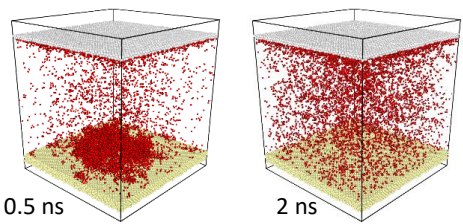

d) Hydrophilic, **out-of-plane**: 100 GHz, **0.5 nm**

**Fig 4** Series of snapshots showing the dynamics of a water droplet on a hydrophilic substrate subjected to in-plane – a,b) and out-of-plane c,d) vibrations at 100 GHz frequency and various amplitudes: a) 0.5 nm, b) 0.75 nm, c) 0.25 nm, d) 0.5 nm.

Three types of processes were distinguished, namely: atomization, boiling, and lift-off. Those processes have been also identified by other authors.[53,54] Atomization is found as a single or few water molecules depart from the water droplet. Boiling, which requires higher energy, has been identified when a collective group of molecules departs from the water droplet. The lift-off process characterizes the entire detachment of the water droplet from the surface

The map of the processes taking place in a hydrophilic system is presented in Figure 3. In the plots, the small circles represent calculations for specific couples of amplitude and frequency values, while the lines and background colors identify the zones where the different phenomena (atomization, boiling, or lift-off) are taken place. The evolution of the droplets with time for the specific representations is shown in Figure 4.

These plots show that the amplitude of the substrate vibrations and not the frequency is the critical variable to activate the water droplet. It is also apparent that, for the same amplitude and frequency, out-of-plane vibrations are more effective than in-plane to activate atomization and boiling phenomena. For large amplitude values, droplet lift-off is observed only for out-of-plane vibrations and not for in-plane ones. This suggests that the energy transfer in the hydrophilic case is more efficient via out-of-plane motion than via in-plane.

The initial stage of water droplet dynamics indicates that the wetting angle progressively decreases with time (see in particular Figure 4b and Figure 4c). This phenomenon has been found experimentally for water droplets activated with SAWs and it is called „acoustowetting effect.[61–63]

For a better comparison of the process efficiencies, Figure 4 shows a series of snapshots of the droplet time evolution for a hydrophilic surface. These snapshots support that for hydrophilic surfaces the atomization/boiling process is much faster for out-of-plane vibrations (compare the evolution of snapshots in Figure 4a and Figure 4d).

Significantly, a different behavior has been observed for hydrophobic surfaces. As observed in Figure 5, the atomization process is in this case less pronounced and the dominant energy transfer mechanism is through droplet lift-off, even for relatively low amplitude values. It can be also realized that the energy transfer induced by out-of-plane vibrations is more efficient than by in-plane ones.

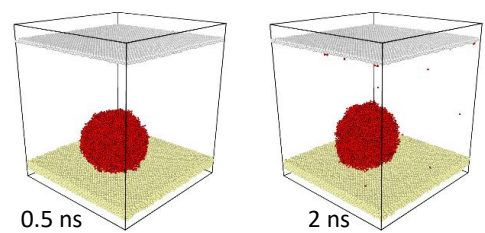

a) Hydrophobic, **in-plane**: 100 GHz, **0.5 nm**

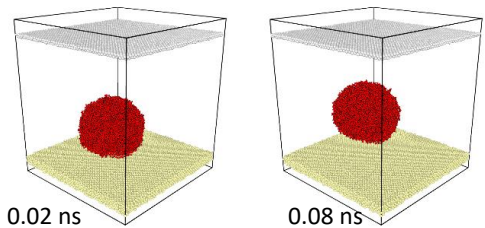

b) Hydrophobic, **in-plane**: 100 GHz, **0.75 nm**

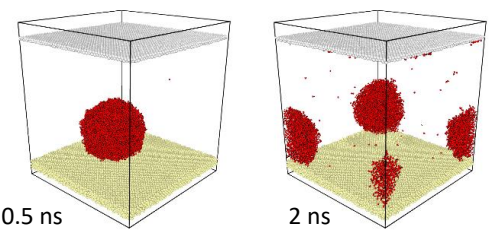

c) Hydrophobic, **out-of-plane**: 100 GHz, **0.25 nm**

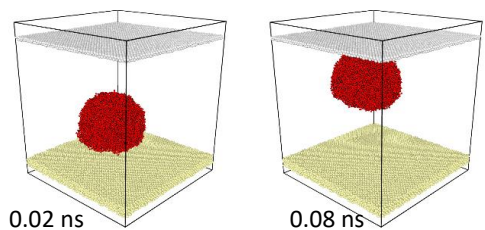

d) Hydrophobic, **out-of-plane**: 100 GHz, **0.5 nm**

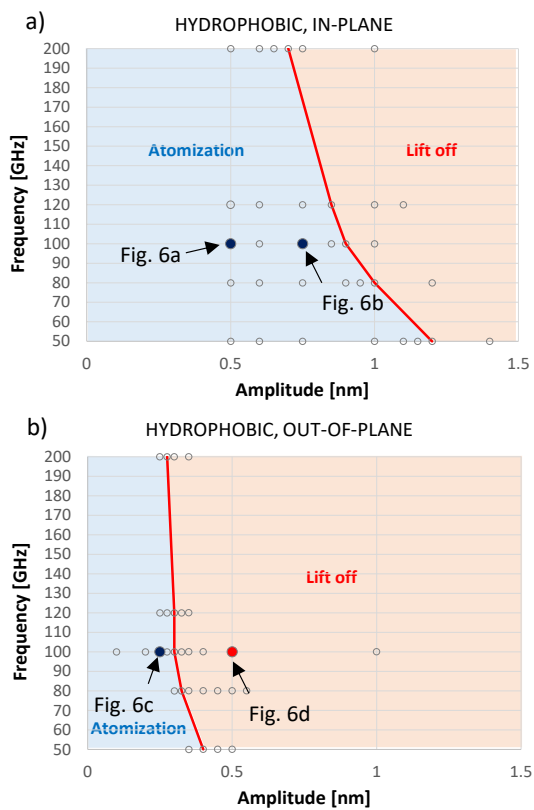

**Fig 5** Maps of excitation phenomena for a water droplet sitting on a superhydrophobic substrate subjected to a) in-plane and b) out-of-plane vibrations as a function of specific amplitude and frequency values.

**Fig 6** Series of snapshots showing the dynamics of a water droplet on a superhydrophobic substrate subjected to in-plane – a,b) and out-of-plane c,d) vibrations at 100 GHz frequency and amplitude a) 0.5 nm, b) 0.75 nm, c) 0.25 nm, d) 0.5 nm.

This is more clearly seen in the series of snapshots for hydrophobic surfaces presented in Figure 6. They show that low energy vibrations (low amplitude and low frequency) give rise to a very slow and limited atomization process (see Figure 6a). Meanwhile, for higher vibration amplitudes, the water droplet starts to move over the surface (see Figure 6b). Application of larger vibrations results in the entire droplet detaching from the surface (lift-off process) (Figures 6b and 6d).

### Vibrational excitation of the ice cluster

Unlike the activation of water droplets with SAWs, where ample literature accounts well for the activation mechanism, only a couple of recent papers[38,39,64] have addressed the analysis of the mechanism of ice activation and its eventual melting.

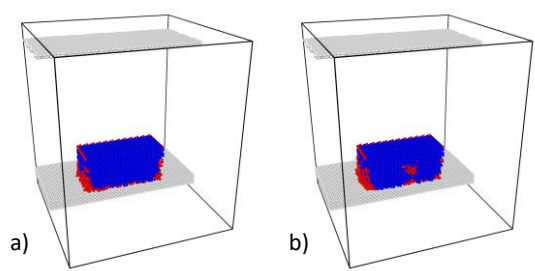

**Fig. 7** The cross-section through the simulated atomic system representing an ice cluster after stabilization for 2ns at 250K on a) hydrophilic and b) hydrophobic surface. Red color – water, Blue color – ice.

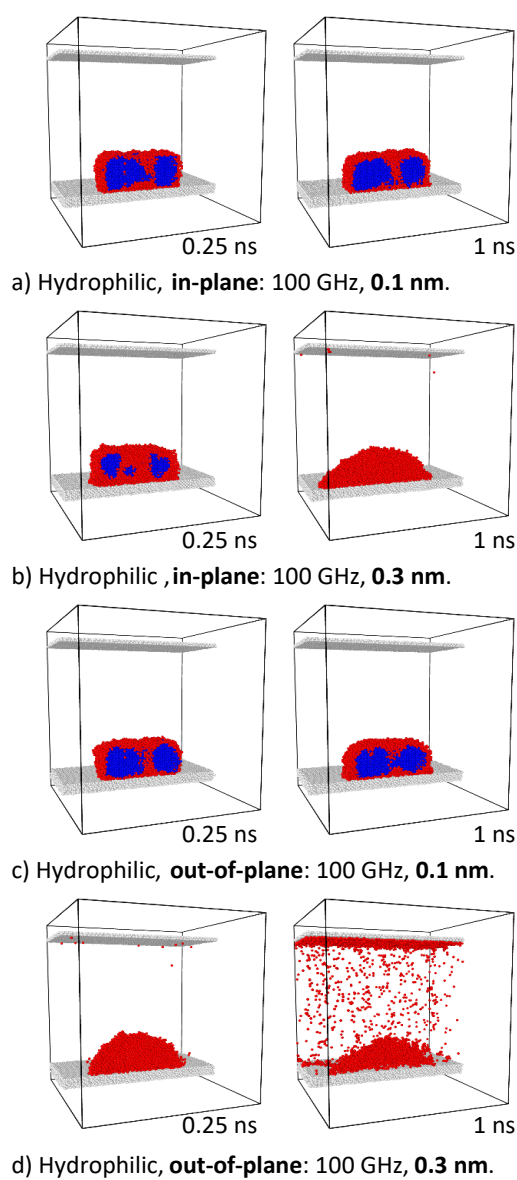

**Fig 8** The dynamics of the ice cluster on the hydrophilic substrate subjected to in-plane – a,b) and out-of-plane c,d) vibrations of 100 GHz frequency and amplitude a,c) 0.1 nm, b,d) 0.3 nm.

The current atomic scale analysis pretends to figure out the parameters and conditions more favorable for the energy transfer from a vibrating substrate to ice. In particular, information will be retrieved on whether excitation via in-plane or out-of-plane vibrations (simulating the effect of normal and shear components of SAWs) are more efficient for de-icing of substrates with a specific wetting behavior (hydrophobic and hydrophilic). In the case of ice clusters, a significantly different response is expected when compared with the behavior of water droplets. Ice contrary to water is difficult to deform plastically and the strain induced by moving the substrate causes relatively large stresses. Those stresses might be slightly reduced when a thin water layer between ice and substrate is present.

Before the analysis of the activation of ice, it is interesting to visualize the initial state of the ice cluster. Figure 7 presents the ice cluster utilized for the simulation after its stabilization at 250 K for 2 ns. This stabilization has been done on the hydrophilic and superhydrophobic surfaces. On the hydrophilic substrate, the reported snapshot reveals the formation of a continuous thin atomic water layer between ice and substrate. On the hydrophobic substrate, some water molecules also form liquid nuclei inside the ice cluster. This has been also observed by other authors.[41,65]

Such an effect is congruent with the fact that a thin water layer appears at the ice/substrate interface. The shear forces induced by in-plane vibrations may easily deform the water layer without transferring the deformation to ice. However, the out-of-plane vibrations cannot be damped by the water atomic layer at the interface and, especially for larger amplitudes, kinetic energy can be effectively transferred to ice.

Starting from these two initial stabilized states at 250 K, the simulation of ice cluster behavior upon activation by the vibrating substrate was carried out utilizing conditions similar to those applied for water droplet activation (see section 3.1.). Thus, in-plane and out-of-plane vibrations were applied to the ice clusters for various frequencies and amplitudes. The obtained molecular dynamics simulation results have shown that ice cluster may start to melt, evaporate or become detached from the surface depending on the energy (i.e., the amplitude of vibration) provided by the vibrating substrate, as well as on its wetting angle. A set of significant results have been gathered in Figures 8 and 9.

The first analysis in Figure 8 refers to the hydrophilic substrate. It has been found that for low amplitude vibrations (0.1 nm – see Figures 8a and 8c) the ice cluster is relatively stable both for in-plane and out-of-plane substrate motions. However, significant differences have been observed for larger amplitudes (0.3 nm – compare Figures 8b and 8d), for which out-of-plane vibrations resulted in a much faster ice cluster melting and subsequent intense water evaporation.

Analysis of ice cluster behavior onto a hydrophobic substrate is shown in Figure 9.

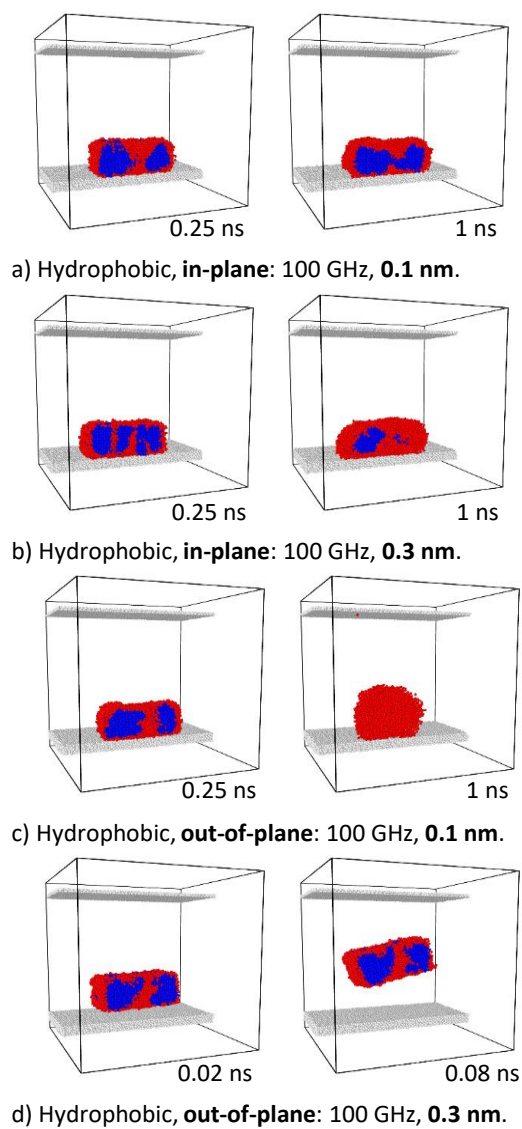

**Fig 9** The dynamics of the ice cluster on the hydrophobic substrate subjected to in-plane – a,b) and out-of-plane c,d) vibrations of 100 GHz frequency and amplitude a,c) 0.1 nm, b,d) 0.3 nm.

Unlike the hydrophilic system, most remarkable in this case is that larger in-plane vibrations do not result in rapid and complete melting of the ice aggregate (compare Figure 8b and Figure 9b) but in partial melting of the outer zones of the ice aggregate where an inner ice core remains after the excitation. This result is in good agreement with the experimental evidence gathered in references,[38,39] the first one dealing with surface acoustic waves and the second one with plate acoustic waves. In both articles, the immediate formation of the liquid water interface between the substrate and the ice has been revealed as part of the de-icing mechanism. Results in Fig. 9 also indicate that in-plane vibrations are less effective for the melting of ice (i.e., to induce a complete de-icing) on hydrophobic than on hydrophilic surfaces.

In further simulations, out-of-plane vibrations were applied to the ice cluster sitting on the hydrophobic substrate (Figure 9). In this case, at low amplitude vibrations, ice melting occurred (see Figure 9c), and for larger amplitudes, the ice cluster lifts off from the surface (see Figure 9d), respectively.

These simulations of the effect of a vibrating substrate on an ice agglomerate support that to attain an effective di-icing (either melting of ice or the detachment of ice from the substrate) surfaces should be preferably hydrophobic and that induced SAW atomic vibrations of the substrate should be tuned to amplify the normal (out-of-plane) wave component. Interestingly, simulations show that for the hydrophilic surfaces the out-of-plane vibrations are also more effective than in-plane, particularly to induce water droplet evaporation (see Figure 4), while than melting of ice is less effective with this excitation mode (Figure 8).

## Summary and Conclusions

Within these studies, atomic scale molecular dynamics simulations were performed to analyze the effectiveness of the kinetic energy transfer from a vibrating substrate to water and ice agglomerates placed on its surface. Numerical modeling has aimed at better understanding the mechanism of kinetic energy transfer from substrate to ice/water. This analysis has provided useful information to design the best conditions for water removal and de-icing systems based on the use of Surface Acoustic Waves (SAW). In de-icing systems based on SAWs, the ratio between a longitudinal and transverse component of Rayleigh wave can be controlled by the elastic properties of the substrate (Poisson ratio), anisotropy of the material as well as the wave actuation conditions. In this regard, the results obtained in the previous analysis have revealed that not only the substrate vibrating modes affect the energy transfer and activation of water and ice. Specifically, we prove that the ice activation modes can be modified by tuning the substrate surface properties towards the reduction of ice adhesion, a characteristic that has resulted indirectly related to the substrate wettability (hydrophobicity or hydrophilicity).

The results of atomic scale simulations performed within these studies have enabled us to identify three phenomena responsible for water removal from the surface, namely: atomization, boiling, and lift-off. However, It has been found that for hydrophilic surfaces lift-off processes are only predictable for a large mechanical energy transfer. Simulations also support that the transverse component of the SAWs (out-of-plane vibrations) would be more efficient for the energy transfer to water, resulting in more efficient energy transfer phenomena as compared with the effect of a longitudinal component (in-plane vibrations). The results also show that water removal mechanisms are different for hydrophilic and hydrophobic surfaces. In the first case, the dominant process is water atomization (evaporation), which transforms into boiling for larger vibration amplitudes. In the hydrophobic case, the water droplet rather rolls over the surface and lifts off for higher vibration energies.

Energy transfer from the vibrating substrate to water and ice is different since liquid accommodates strain by plastic deformation and the ice is not easily deformed. The results of molecular dynamics simulations indicate that both in-plane and out-of-plane vibrations are more effective in the case of ice. This effect is confirmed by the molecular dynamics studies performed herewith. It can be found that in-plane vibrations are much less effective than out-of-plane, especially for larger vibration amplitudes. Larger amplitude in-plane vibrations result in ice melting for the hydrophilic surface, which is not the case for the hydrophobic one. On the other hand, for out-of-plane vibrations, the ice on the hydrophobic surface melts at lower amplitude, while for larger amplitude rapidly lift-off the surface.

The above-mentioned results from our simulation analysis can be translated into some practical guidelines for the experimental design of an active, SAW-based, de-icing system. In particular:

1. The selection of the surface state as well as the design of SAW actuators should be oriented to amplify the out-of-plane wave component of the vibration since it is more effective for both anti-icing and de-icing processes.
2. The hydrophobic state of the surface facilitates the melting of ice even for relatively low amplitudes of out-of-plane vibrations. For hydrophobic surfaces, the in-plane vibrations are insufficient to cause the melting of the whole ice cluster.
3. Hydrophilic surfaces and large amplitudes facilitate the melting of ice for the in-plane component of SAW when larger amplitudes are applied.

## Author Contributions

T.W. Investigation, methodology, formal analysis, writing – original draft, writing – review & editing, funding acquisition. S.J. Conceptualization, formal analysis, writing – review & editing. A.W. Conceptualization, writing – review & editing, supervision, funding acquisition. A.R.G-E. Writing – review & editing, supervision, funding acquisition. A.B. Conceptualization, writing – review & editing, funding acquisition.

## Conflicts of interest

There are no conflicts to declare.

## Acknowledgements

The project leading to this article has received funding from the EU H2020 program under grant agreement 899352 (FETOPEN-01-2018-2019-2020 - SOUNDofICE).